\documentclass[11pt,letterpaper]{article}

\usepackage{units}

\usepackage{natbib}
\usepackage{graphicx}
\usepackage{amsmath}
\usepackage{amssymb}
\usepackage{afterpage}
\usepackage{graphics}
\usepackage{float}
\usepackage{lineno}
\usepackage{rotating}
\usepackage{xspace}
\usepackage{dcolumn} 
\usepackage{bm} 

\usepackage[left=1.0in,top=1.0in,bottom=1.0in,right=1.0in]{geometry}

\title{\textbf{Probing the accelerating Universe with radio weak lensing in the JVLA Sky Survey}}
\date{} 

\newcommand{\be}{\begin{equation}}
\newcommand{\ee}{\end{equation}}

\begin{document}

\vspace{-3.5cm}
\maketitle
\renewcommand*{\thefootnote}{\fnsymbol{footnote}}
\vspace{-2.0cm}
\begin{center}
M.~L.~Brown$^{1}$\footnote{Corresponding author: m.l.brown@manchester.ac.uk}, 
F.~B.~Abdalla$^{2}$, 
A.~Amara$^{3}$, 
D.~J.~Bacon$^{4}$,
R.~A.~Battye$^{1}$, 
M.~R.~Bell$^{5}$, 
R.~J.~Beswick$^{1}$, 
M.~Birkinshaw$^{6}$,
V.~B\"ohm$^{5}$, 
S.~Bridle$^{1}$, 
I.~W.~A.~Browne$^{1}$, 
C.~M.~Casey$^{7}$, 
C.~Demetroullas$^{1}$, 
T.~En\ss lin$^{5,8}$,
P.~G.~Ferreira$^{9}$, 
S.~T.~Garrington$^{1}$, 
K.~J.~B.~Grainge$^{1}$,
M.~E.~Gray$^{10}$, 
C.~A.~Hales$^{11}$, 
I.~Harrison$^{1}$,
A.~F.~Heavens$^{12}$,
C.~Heymans$^{13}$, 
C.-L.~Hung$^{14}$, 
N.~J.~Jackson$^{1}$, 
M.~J.~Jarvis$^{9}$, 
B.~Joachimi$^{2}$, 
S.~T.~Kay$^{1}$, 
T.~D.~Kitching$^{15}$, 
J.~P.~Leahy$^{1}$, 
R.~Maartens$^{16,4}$,
L.~Miller$^{9}$, 
T.~W.~B.~Muxlow$^{1}$, 
S.~T.~Myers$^{17}$, 
R.~C.~Nichol$^{4}$,
P.~Patel$^{16,18}$, 
J.~R.~Pritchard$^{12}$, 
A.~Raccanelli$^{19,20}$, 
A.~Refregier$^{3}$, 
A.~M.~S.~Richards$^{1}$, 
C.~Riseley$^{21}$, 
M.~G.~Santos$^{16,22}$,
A.~M.~M.~Scaife$^{21}$, 
B.~M.~Sch\"afer$^{23}$, 
R.~T.~Schilizzi$^{1}$,
I.~Smail$^{24}$,
J.-L.~Starck$^{25}$, 
R.~M.~Szepietowski$^{1}$, 
A.~N.~Taylor$^{13}$, 
L.~Whittaker$^{1}$,
N.~Wrigley$^{1}$,
J.~Zuntz$^{1}$\\
\vspace{0.5cm}
\emph{(Affiliations can be found after the references)}

\end{center}
\renewcommand*{\thefootnote}{\arabic{footnote}}
\setcounter{footnote}{0}
\normalsize
\vspace{-0.50cm}
\begin{abstract}
\bf{We outline the prospects for performing pioneering radio weak
  gravitational lensing analyses using observations from a potential
  forthcoming JVLA Sky Survey program. A large-scale survey with the
  JVLA can offer interesting and unique opportunities for performing
  weak lensing studies in the radio band, a field which has until now
  been the preserve of optical telescopes. In particular, the JVLA has
  the capacity for large, deep radio surveys with relatively high
  angular resolution, which are the key characteristics required for a
  successful weak lensing study. We highlight the potential advantages
  and unique aspects of performing weak lensing in the radio band. In
  particular, the inclusion of continuum polarisation information can
  greatly reduce noise in weak lensing reconstructions and can also
  remove the effects of intrinsic galaxy alignments, the key
  astrophysical systematic effect that limits weak lensing at
  all wavelengths. We identify a VLASS ``deep fields'' program (total
  area $\sim$10--20 deg$^2$), to be conducted at L-band and with
  high-resolution (A-array configuration), as the optimal survey
  strategy from the point of view of weak lensing science. Such a
  survey will build on the unique strengths of the JVLA and will
  remain unsurpassed in terms of its combination of resolution and
  sensitivity until the advent of the Square Kilometre Array. We
  identify the best fields on the JVLA-accessible sky from the point
  of view of overlapping with existing deep optical and near infra-red
  data which will provide crucial redshift information and facilitate
  a host of additional compelling multi-wavelength science.}
\end{abstract}

\section{Introduction}
\label{intro}
Weak gravitational lensing is the effect whereby images of faint
and distant background galaxies are coherently distorted due to
deflection of their light by intervening large scale structures in the
Universe. This ``cosmic shear'' effect is recognised as one of the key cosmological probes that
will allow us to precisely probe the nature of dark energy
with future surveys~\citep{albrecht06, peacock06}. The current
state-of-the-art in terms of weak lensing comes from
optical surveys covering 154 deg$^2$, i.e.~the recent CFHTLenS
results \citep{heymans12}. Although the lensing-derived constraints on
the evolution of structure are currently not strong enough to
meaningfully constrain the properties of dark energy, ongoing and
future ground-based surveys, e.g.~the KiDS~\citep{deJong13}, 
DES\footnote{http://www.darkenergysurvey.org},
HSC\footnote{http://www.naoj.org/Projects/HSC/}, LSST\footnote{https://www.lsstcorp.org} and
SKA\footnote{https://www.skatelescope.org} surveys, and ultimately satellite
missions such as NASA's {\it
  WFIRST}\,\footnote{http://wfirst.gsfc.nasa.gov} and ESA's {\it
  Euclid}\,\footnote{http://sci.esa.int/euclid/} telescope~\citep{laureijs11}, promise to revolutionise the field of weak lensing by allowing
precision measurements of structure growth. In addition, lensing
measurements can be used to test the nature of gravity in a
complementary way to other cosmological probes
(e.g.~\citealt{simpson13, raccanelli12}). 

The only detection of weak lensing in the radio band to date was made by
\cite{chang04} using the VLA FIRST
survey~\citep{becker95}. Since then progress in radio weak lensing
studies has lagged behind the optical because of the much smaller
number density of galaxies typically seen in radio surveys as compared
to the optical bands. This situation is beginning to change with the
advent of a new generation of radio telescopes. Indeed a number of
relatively large observational programs in the radio have weak lensing
as one of their primary science drivers. In particular the
SuperCLASS\footnote{http://www.e-merlin.ac.uk/legacy/projects/superclass.html}
survey on the UK's e-MERLIN telescope aims to detect the weak lensing
signal in a supercluster of galaxies while the
CHILES\footnote{http://www.mpia-hd.mpg.de/homes/kreckel/CHILES/index.html}
continuum and HI surveys, currently being undertaken on the JVLA, will
search for radio weak lensing effects in the COSMOS field. Large
scale surveys with the LOFAR telescope and with the SKA pathfinder
telescopes, MeerKAT and ASKAP will also offer interesting
opportunities for radio weak lensing studies (mainly through
lensing magnification effects) in the run-up to Phase-1 of the SKA
for which construction is due to start in 2017. 

This new generation of radio telescopes can offer unique and powerful
added value to the field of weak lensing. Firstly, deep radio surveys
will probe the lensing power spectrum at significantly higher redshift
than most of the planned optical lensing surveys. The addition of
radio can therefore offer a more powerful redshift ``lever arm'' with
which to measure the effects of dark energy on the evolution of
structure. Secondly, instrumental systematic effects are a serious
concern for weak lensing studies for which a very accurate
representation of the beam or point spread function (PSF) of the
telescope is required. The highly stable and deterministic beam
response of radio interferometers could therefore prove a major
advantage for weak lensing science. Thirdly, the radio offers unique
and novel opportunities to measure the effects of weak lensing that
are not available to optical lensing surveys through polarization
measurements, HI rotational velocity measurements and the direct
measurement of galaxy shapes in the $uv$ visibility plane. 

The JVLA's unique combination of excellent sensitivity and relatively
high angular resolution will remain unsurpassed until the advent of
the SKA. These qualities also make the JVLA an excellent facility with
which to spearhead the development of radio weak lensing. Here, we
describe how the JVLA could play a major role in this rapidly
developing field through consideration of radio weak lensing science
during the survey design for a new generation of VLA Sky Surveys
(VLASS).

\section{Optimal survey configurations for the VLASS}
\label{survey_optimize}
Here we examine the optimal configurations for a VLASS
conducted at frequencies 1.4 (L-band), 3.0 (S-band) and 4.8~GHz
(C-band). Using Fisher analyses and simple mode-counting arguments,
one can predict the achievable errors on the cosmic shear power
spectrum with a given survey design. For our purposes, by ``survey
design'', we simply mean the survey area and depth (the latter assumed
to be constant across the survey). The forecasted errors on a
measurement of the 2D cosmic shear power spectrum in a band of $\Delta
\ell$, centred on a multipole, $\ell$ are
\be \Delta C_\ell =
\sqrt{\frac{2}{(2\ell + 1) \Delta\ell f_{\rm sky}}} \left[ C_{\ell} +
  \frac{\gamma^2_{\rm rms}}{2 n_g} \right], 
\label{eq:cl_err}
\ee 
where $f_{\rm sky}$ is the fraction of sky observed, $n_g$ is the
number density of source galaxies and $\gamma_{\rm rms}$ is the total
dispersion in the galaxy ellipticity estimates due to both measurement
errors and the intrinsic dispersion in galaxy shapes. We assume both
to be $\sim 0.3$ resulting in an effective shear estimator dispersion
of $\gamma_{\rm rms} = 0.42$. This is consistent with the observed
shapes of galaxies in both deep optical surveys
(e.g.~\citealt{miller13}) and in deep radio
observations~\citep{patel10}.

$C_\ell$ is the power spectrum which we calculate in the concordance
$\Lambda$CDM cosmology as
\be
C_{\ell} = \int_0^{\chi_h} \frac{W^2(\chi)}{\chi^2} P_{\delta}(\ell/\chi, \chi) \, d\chi 
\ee
where $\chi_h$ is the comoving distance to the horizon. $W(\chi)$
is the lensing efficiency function, which for a lens at comoving distance,
$\chi_d$ and redshift, $z_d$ is
\be
W(\chi_d) = \frac{3 H_0^2 \Omega_m}{2 c^2} \chi_d (1 + z_d)
\int_{\chi_d}^{\chi_h} f(\chi_s) \frac{(\chi_s - \chi_d)}{\chi_s} \, d\chi_s.
\ee 
Here, $f(\chi)$ is the distribution of sources in comoving distance,
specified by the source redshift distribution, $f(\chi)d\,\chi =
n(z)\,dz$. Note that, in this work, we have considered only the measurement of a single
projected 2D power spectrum ($C_\ell$). However, more generally, given
distance information (e.g.~from photometric redshift estimates), one
can extract additional cosmological information on the
evolution of structure in the Universe by measuring the lensing power
spectrum in a series of tomographic redshift bins, $C_\ell(z)$.

\begin{figure}[t!]
\vspace{-10mm}
\centering
\includegraphics[width=12cm]{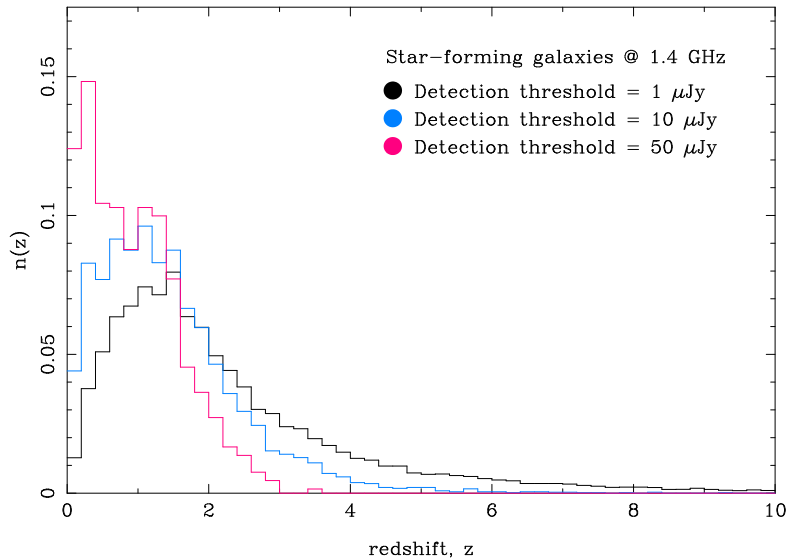}
\vspace{-5mm}
\caption{Redshift distribution of star-forming galaxies in the SKADS
  simulation for L-band detection thresholds of 1, 10 and 50
  $\mu$Jy. The distributions have median redshifts of $1.76, 1.25$ and
  $0.84$ respectively and are normalized arbitrarily such that $\sum_z
  n(z) = 1$. For the same detection threshold, the S-band and
  C-band distributions are broadly similar to the L-band $n(z)$ although
  the total number of galaxies detected drops significantly
  (Fig.~\ref{fig:gal_densities_sizes}).}
\label{fig:nofz_dist}
\end{figure}

To calculate error forecasts for different survey designs, we
require the redshift distribution of sources as a function
of flux density threshold. Throughout this work we adopt a detection
threshold of $S_{\rm tot} > 10\sigma$ where $\sigma$ is the projected
RMS image noise. We have obtained an estimate of the radio-frequency
$n(z)$ from the Square Kilometre Array Design Studies (SKADS)
simulation~\citep{wilman08} for flux density thresholds between $1$ and $100 \,
\mu$Jy. Fig.~\ref{fig:nofz_dist} shows how the normalized 1.4 GHz
$n(z)$ changes with the detection threshold for only the star-forming
galaxies included in the SKADS simulation. For similar detection
thresholds the redshift distributions for the S-band and C-band galaxy
populations in the simulation are broadly similar.

\begin{figure}[t!]
\vspace{5mm}
\centering
\includegraphics[width=15cm]{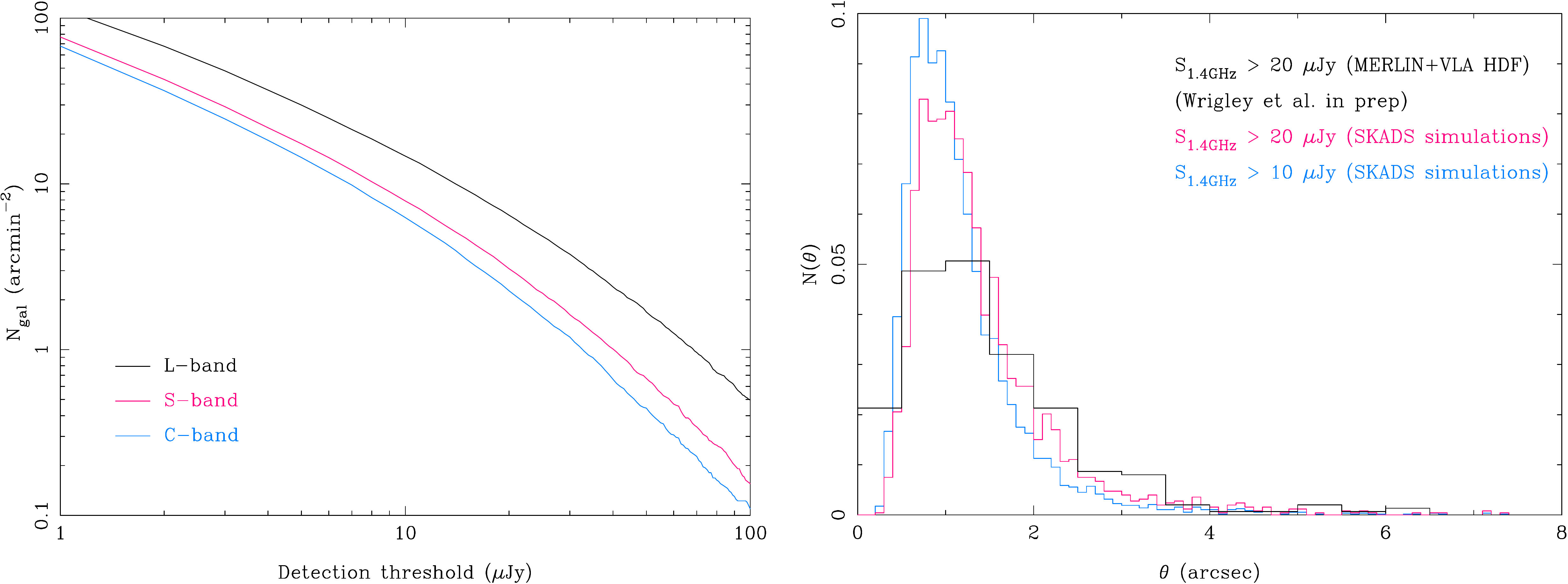}
\vspace{1mm}
\caption{\emph{Left panel:} The projected surface number density of
  galaxies as a function of the detection threshold for the JVLA's L-,
  S- and C-band frequency ranges. These projections have been
  obtained from the SKADS simulation, renormalized to match the
  observed counts of $N(S_{\rm 1.4 \, GHz} \gtrsim 50 \, \mu\rm Jy)
  \approx 1.5$ arcmin$^{-2}$ as measured in deep existing radio
  observations such as the VLA + MERLIN observations of the
  HDF-North~\protect\citep{muxlow05} and the Large and Deep components
  of the VLA-COSMOS survey~\protect\citep{schinnerer10}. \emph{Right
    panel:} Distribution of galaxy sizes for limiting flux densities
  of $S_{\rm 1.4 \, GHz} = 10$ and $20$ $\mu$Jy as measured from the SKADS
  simulation and re-normalized (reduced in size by a factor of
  $\sim$ 3) to agree with the typical sizes of galaxies detected in
  the HDF-North. Also plotted are the actual measured sizes of
  galaxies as seen in a re-analysis of the MERLIN + VLA HDF data
  making use of size estimates for 339 galaxies at $S_{\rm 1.4 \, GHz}
  > 20\,\mu$Jy (Wrigley et al.\!\!\!~, in prep.). For most of the survey
  configurations we examine in this study, the majority of
  detected galaxies will be resolved by the VLASS and so will
  allow accurate shape measurements.}
\label{fig:gal_densities_sizes}
\end{figure}

The SKADS simulation also provides us with an estimate of the galaxy
surface number density for a given detection threshold. These are
plotted in the left panel of Fig.~\ref{fig:gal_densities_sizes} for
the three VLA frequency bands considered here. Note that we have
re-normalized the galaxy densities such that $N(S_{1.4 \, \rm GHz} >
54 \, \mu \rm Jy) \approx 1.5$ arcmin$^{-2}$ in order to match the number
counts seen in the deep radio surveys of the VLA + MERLIN HDF-North
field~\citep{muxlow05} and the VLA-COSMOS Large and Deep
surveys~\citep{schinnerer10}. Our adopted number density
normalization is mid-way between those of the ``gold'' ($N(S_{1.4 \, \rm GHz} >
54 \, \mu \rm Jy) = 0.75$ arcmin$^{-2}$) and ``silver'' ($N(S_{1.4 \, \rm
  GHz} > 54 \, \mu \rm Jy) = 3.76$ arcmin$^{-2}$) galaxy samples
constructed by \cite{patel10} who re-analysed the \cite{muxlow05} VLA
+ MERLIN data for the purposes of a radio weak lensing study.

A further crucial factor that can impact the performance of weak
lensing studies is the angular resolution of the telescope in relation
to the typical size of the galaxies for which one wishes to obtain
accurate shape measurements. In the right panel of
Fig.~\ref{fig:gal_densities_sizes}, we plot the distribution of galaxy
sizes for two representative detection thresholds (10 and 20
$\mu$Jy). Once again, these distributions have been derived from the
SKADS simulation but they have been re-normalized to agree with
observations. This re-normalization (which required a reduction in the
sizes listed in the SKADS simulation by a factor of $\sim$3) was
necessary to bring the mean galaxy size for the $S_{\rm 1.4 \, GHz} > 20$
$\mu$Jy sample down to around 1 arcsec as found in a re-analysis of
the \cite{muxlow05} MERLIN + VLA HDF-North observations (Wrigley et
al.\!\!\!,~in prep.). For well-detected
(e.g. $\gtrsim 10\sigma$) galaxies it is reasonable to assume that
a good shape measurement is possible if the galaxy size is more than
about 50\% of the angular resolution of the telescope. Inspection of
Fig.~\ref{fig:gal_densities_sizes} then suggests a typical required
angular resolution of $\sim$1 arcsec for a detection threshold
of 10--20 $\mu$Jy. Comparing to the resolution
capabilities of the JVLA, the optimal array configuration would likely
be A-array for L- and S-band observations (1.3 and 0.65 arcsec
resolution respectively) while B-array would be appropriate for C-band
observations (1.0 arcsec resolution). We note that for observations at
S-band in A-array configuration, one would also need to consider how
much of a typical galaxy's large scale emission is resolved out and
the resulting potential impact on required survey times. This should
not be a problem at L-band for which the A-array resolution is
well-matched to typical galaxy sizes. 

To arrive at the optimal survey strategy for detecting cosmic
shear for a given amount of telescope time, we keep the quantity
$\sqrt{\Omega} / S_{\rm rms}$ fixed where $\Omega$ is the survey area
and $S_{\rm rms}$ is the RMS noise in flux density. This relation is
normalized using the survey speed parameters required to achieve 100
$\mu$Jy image noise RMS as listed in the VLASS capabilities
document\footnote{see https://science.nrao.edu/science/surveys/vlass/capabilities}
made available alongside the call for VLASS White papers.

\begin{figure}[t!]
\vspace{5mm}
\centering
\includegraphics[width=16cm]{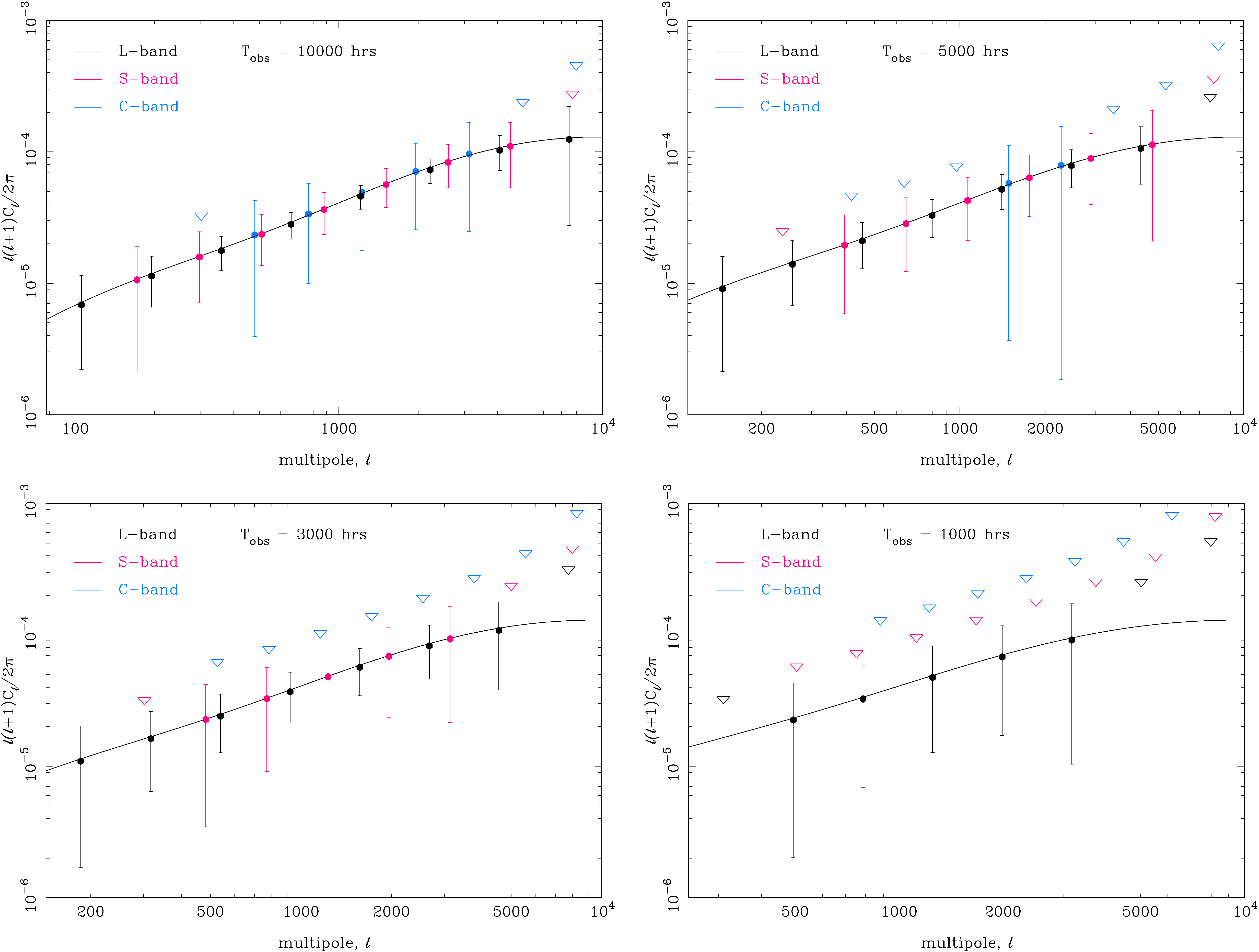}
\vspace{1mm}
\caption{Forecasted constraints on the weak lensing power spectrum for
  the optimal L-, S- and C-band survey configurations for observing times of
  10000 (\emph{upper left}), 5000 (\emph{upper right}), 3000
  (\emph{lower left}) and 1000 (\emph{lower right}) hours. Open wedges
  indicate upper limits.}
\label{fig:cls_combined}
\end{figure}

The signal-to-noise of the detection of cosmic shear is calculated as
\be 
S/N = \left(\sum_b P_b^2 / \sigma_b^2\right)^{1/2}, 
\ee 
where the sum is over shear power spectrum bandpowers ($P_b$) and
$\sigma_b$ is the forecasted error on each bandpower
(eq.~\ref{eq:cl_err}). We calculate $S/N$ as a function of the RMS
noise flux density and survey area. Note that the minimum and maximum
observable power spectrum multipoles are also determined by the survey
configuration through the maximum survey dimension and the typical
angular separation of nearest neighbour galaxies respectively. These
effects are included in our power spectrum forecasts. Finally, the
optimal survey configuration is chosen to be the one which maximizes
the $S/N$. We have performed this survey optimization procedure for four cases of
total observation time, $T_{\rm obs} = 1000, 3000, 5000$, and $10000$
hours. The predicted constraints on the shear power spectrum for the
optimal survey found in each case are shown in Fig.~\ref{fig:cls_combined}.

\begin{table*}
\centering
\caption{Optimal survey areas and shear power spectrum detection
  significances for different observing times at L-band, S-band and
  C-band. The first column lists the assumed observation time in hours.
  The following six columns list the optimal survey area in square
  degrees ($\Omega$) and the corresponding signal-to-noise of the
  weak lensing power spectrum detection ($C_\ell^{\rm lens}$ S/N) for the
  three frequency bands. The required depths (in terms of image
  RMS) are $\sim1\,\mu$Jy, $0.6\,\mu$Jy and $0.5\,\mu$Jy for L-, S-
and C-band respectively, independent of observation
time.} \label{table1}
\vspace{0.3cm}
\begin{tabular}{ccccccc}
\noalign{\vskip 3pt\hrule\vskip 5pt}
$T_{\rm obs}$ (hrs) & $\Omega$ (deg$^2$) & $C_\ell^{\rm lens}$ S/N & $\Omega$ (deg$^2$) & $C_\ell^{\rm lens}$ S/N & $\Omega$ (deg$^2$) & $C_\ell^{\rm lens}$ S/N \\
      & (L-band) & (L-band) & (S-band) & (S-band) & (C-band) & (C-band)\\
\noalign{\vskip 3pt\hrule\vskip 5pt}
1000   &   1.7 & 3.0 & 0.6 & 1.9 & 0.2 & 0.9 \\
3000   &   5.0 & 5.4 & 1.8 & 3.4 & 0.5 & 1.8 \\
5000   &   8.4 & 7.0 & 3.0 & 4.4 & 0.9 & 2.4 \\
10000  &  16.8 & 9.9 & 6.0 & 6.3 & 1.8 & 3.5 \\
\noalign{\vskip 5pt\hrule\vskip 3pt}
\end{tabular}
\footnotetext[1]{table footnote 1}
\end{table*} 

Table~\ref{table1} lists the survey area and $S/N$ values for the
optimal survey for each $T_{\rm obs}$ and frequency band
considered. We see that for the largest survey duration of 
$T_{\rm obs} = 10000$ hrs, we could expect to detect the cosmic shear
signal at $>\!\!3\sigma$ even if the survey were to be conducted at
C-band (although one would have to spend all of that time integrating on
only 2 square degrees of sky). At L-band the $T_{\rm obs} = 10000$
hours survey could lead to a $\sim\!10\sigma$ detection which is
approaching the sensitivity of the most sensitive optical lensing
surveys conducted to date. Note that to achieve this $10\sigma$
detection, once again, one would need to concentrate on a relatively
small field ($\sim\!20$ square degrees). 

\begin{figure}[t!]
\vspace{5mm}
\centering
\includegraphics[width=15cm]{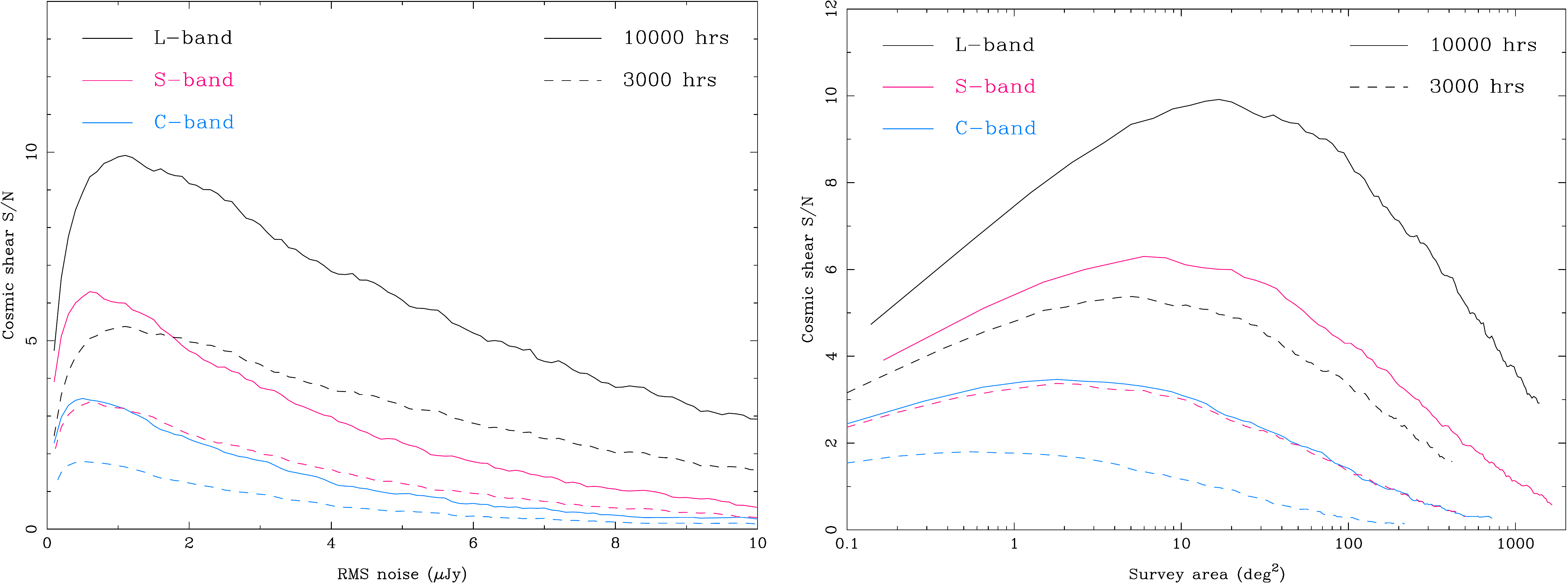}
\vspace{1mm}
\caption{The signal-to-noise with which the shear power spectrum is
  detected for the 3000 and 10000 hours survey in each frequency
  band as a function of RMS noise (\emph{left panel}) and survey area
  (\emph{right panel}). The corresponding plots for the other survey times
  investigated look broadly similar with the optimal survey areas
  increasing with increasing observation time.}
\label{fig:opt_curves}
\end{figure}

It is interesting to examine how the $S/N$ of the detection depends on
the adopted RMS image noise and survey size. These dependencies are
demonstrated in Fig.~\ref{fig:opt_curves} where we plot the $S/N$
curves for each frequency band for the $T_{\rm obs} = 10000$ and
$T_{\rm obs} = 3000$ hrs cases. The left hand panel of
Fig.~\ref{fig:opt_curves} shows $S/N$ as a function of RMS noise. We
see that the optimal image noise RMS level is insensitive to the
survey time adopted, remaining at $\sim\!1$ $\mu$Jy for the L-band
survey for both values of $T_{\rm obs}$. Similarly for the S- and
C-band surveys, the preferred image RMS noise levels are $\sim\!0.6$
$\mu$Jy and $\sim\!0.5$ $\mu$Jy respectively independent of survey
time. The corresponding preferred galaxy number density in all three
cases is $\sim15$ arcmin$^{-2}$. Note that these conclusions would
change to some degree if we were to change the measure which we wish
to optimize. For example, if we were to optimize on the survey's
ability to constrain the very large scale power spectrum (for instance
only multipoles $\ell < 100$), then a higher noise level would be
preferred as the relative importance of the sample variance and noise
variance terms in eq.~(\ref{eq:cl_err}) would have been altered.

The right hand panel of Fig.~\ref{fig:opt_curves} shows the $S/N$ as a
function of survey area again for the $T_{\rm obs} = 10000$ and
$T_{\rm obs} = 3000$ hrs cases. As expected, we see the preferred
survey areas increasing as we increase the observation time. Perhaps
the most useful aspect of this plot is the ability to identify the
range of survey areas over which the $S/N$ remains approximately
constant for a given survey duration. It is clear from the figure that
the $S/N$ curves become more peaked for longer and deeper surveys
meaning that this range is smaller for larger survey
programs. Nevertheless, even for the most powerful radio weak lensing
survey (10000 hrs @ L-band), the $S/N$ is relatively
constant for survey areas between $\sim 10$ and $\sim 100$ square
degrees. For ease of comparison, in Table~\ref{table2}, we report the
$S/N$ values for all of the survey configurations that we have
investigated for the cases of $\Omega = 1, 10$ and $100$ square
degrees. 

\begin{table*}
\centering
\caption{Detection significances for the case of JVLA surveys covering 1,
  10 and 100 deg$^2$ for different observing times, $T_{\rm
    obs}$ and for each of the L-, S-, and C-band channels.}\label{table2}
\vspace{0.3cm}
\begin{tabular}{rcccc}
\noalign{\vskip 3pt\hrule\vskip 5pt}

       & $C_\ell^{\rm lens}$ S/N  & $C_\ell^{\rm lens}$ S/N & $C_\ell^{\rm lens}$ S/N \\
       & $\Omega = 1$ deg$^2$ & $\Omega = 10$ deg$^2$ & $\Omega = 100$ deg$^2$ \\

\noalign{\vskip 3pt\hrule\vskip 5pt}
10000 hrs @ L-band  & 6.9 &  9.2  & 8.4  \\
5000  hrs @ L-band  & 5.6 &  6.9  & 5.0  \\
3000  hrs @ L-band  & 4.7 &  5.0  & 3.4  \\
1000  hrs @ L-band  & 3.1 &  2.6  & 1.1  \\
\\
10000 hrs @ S-band  & 4.7 &  5.9  & 4.2  \\
5000  hrs @ S-band  & 3.6 &  4.2  & 2.3  \\
3000  hrs @ S-band  & 2.9 &  2.8  & 1.4  \\
1000  hrs @ S-band  & 1.7 &  1.2  & 0.3  \\
\\
10000 hrs @ C-band  & 3.2 &  3.0  & 1.4  \\
5000  hrs @ C-band  & 2.0 &  1.7  & 0.6  \\
3000  hrs @ C-band  & 1.7 &  1.1  & 0.3  \\
1000  hrs @ C-band  & 0.8 &  0.4  & 0.1  \\
\noalign{\vskip 5pt\hrule\vskip 3pt}
\end{tabular}
\footnotetext[1]{table footnote 2}
\end{table*} 

\section{The unique value of weak lensing in the radio}
\label{novel_techniques}
One of the most compelling reasons for pursuing measurements of weak
lensing in the radio band is the unique added value that radio lensing
can offer above and beyond the traditional optical-based
approaches. One unique advantage is the polarization information
that is available in the radio and which can provide information on
the intrinsic (unlensed) shapes of background galaxies. As described in
\cite{brown11a}, the position angle of the integrated polarized
emission from a background galaxy is unaffected by gravitational
lensing. If the polarized emission (which is polarized synchrotron
emission sourced by the galaxy's magnetic field) is also strongly
correlated with the disk structure of the galaxy then measurements of
the radio polarization position angle can be used as estimates of the
galaxy's intrinsic (unlensed) position angle. 

Such an approach could potentially have two key advantages over
traditional weak lensing analyses. Firstly, the polarization technique
can be used to effectively remove the primary astrophysical
contaminant of weak lensing measurements -- intrinsic galaxy
alignments (see e.g.~\citealt{heavens00, catelan01, hirata04, brown02})
-- which are a severe worry for ongoing and future precision cosmology
experiments based on weak lensing. Secondly, depending
on the polarization properties of distant background disk galaxies,
the polarization technique has the potential to reduce the effects of
noise due to the intrinsic dispersion in galaxy shapes. Using the
polarization technique, the forecasted errors on a measurement of the
shear power spectrum becomes~\citep{brown11b}
\be \Delta C_\ell =
\sqrt{\frac{2}{(2\ell + 1) \Delta\ell f_{\rm sky}}} \left[ C_{\ell} +
  \frac{16 \alpha^2_{\rm rms}\gamma^2_{\rm rms}}{2 n_{\rm pol}} \right], 
\label{eq:cl_err_pol}
\ee 
where $\alpha_{\rm rms}$ is the scatter in the relationship between
the observed polarization position angle and the intrinsic structural
position angle of the galaxy and $n_{\rm pol} \approx f_{\rm pol} n_g$ is
the number of galaxies for which one can obtain accurate polarization
measurements. These parameters depend on the details of the
polarization properties of background galaxies (e.g.\!~the mean
polarization fraction, $\Pi_{\rm pol}$) which are currently not well
known. There are some existing measurements for a sample of local spiral 
galaxies~\citep{stil09} which suggest $\alpha_{\rm rms} < 15^\circ$
and $\Pi_{\rm pol} < 20\%$ although the sample is small. 

\begin{figure}[t!]
\vspace{-10mm}
\centering
\includegraphics[width=12cm]{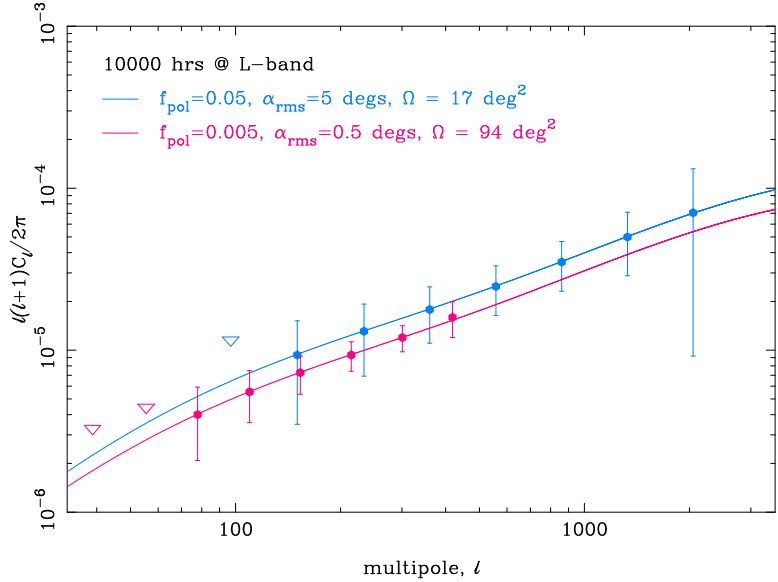}
\vspace{-5mm}
\caption{Forecasts of the constraints on the shear power spectrum
  obtainable with a 10000 hour L-band survey using the polarization
  technique described in the text. The observational parameters,
  $f_{\rm pol}$ and $\alpha_{\rm rms}$, appropriate for the faint
  $\mu$Jy galaxy population are unknown at this time so we present
  forecasts for two possible cases. The forecasted constraints are
  restricted to the low-$\ell$ (linear) part of the power spectrum due
  to the assumed low surface number density of galaxies with
  measurable polarization.}
\label{fig:pol_forecasts}
\end{figure}

In Fig.~\ref{fig:pol_forecasts}, we plot the forecasted constraints
that could be achieved with a 10000 hr L-band survey for two
representative cases: $\{\alpha_{\rm rms} = 5^\circ; f_{\rm pol} =
5\%\}$ and $\{\alpha_{\rm rms} = 0.5^\circ; f_{\rm pol} =
0.5\%\}$. Once again, in each case we plot the forecasts for the
optimal survey area and depth which was identified as described in the
previous section. It is clear from eq.~(\ref{eq:cl_err_pol}) that
the polarization technique becomes free of shape noise in the limit of
$\alpha_{\rm rms} = 0$. A consequence of this behaviour is apparent
in Fig.~\ref{fig:pol_forecasts} -- as $\alpha_{\rm rms}$ is reduced
the shape noise contribution to the total error becomes sub-dominant
and the optimization procedure pushes the survey area to larger
areas. At the same time, a reduction in the mean polarization
fraction, $\Pi_{\rm pol}$ will result in a low surface number density
of galaxies which will limit the maximum multipole that can be probed
with the polarization technique. Note also that it may be possible
to select sub-samples of the total galaxy population to have
particular polarization properties. For example, one could imagine
that selecting only galaxies with high fractional polarization would
yield a galaxy sub-sample with highly ordered magnetic fields which
would consequently have a very tight correlation (low $\alpha_{\rm
  rms}$) between the polarization orientations and the intrinsic
structural position angles of the galaxies. Of course, such a
sub-sample would also have a very low surface number density of
galaxies. The polarization technique may therefore be better suited to
probing the shear power spectrum on large scales where high number
densities are not required. 

A second novel idea that is well suited to radio observations is to
use rotational velocity measurements to provide information about the
intrinsic shapes of galaxies. The idea, first suggested by
\cite{blain02} and \cite{Morales06}, is to measure the axis of
rotation of a disk galaxy and to compare this to the orientation of
the major axis of the galaxy disk image. In the absence of lensing,
these two orientations should be perpendicular and measuring the
departure from perpendicularity directly estimates the shear field at
the galaxy's position. Such an analysis would require commensal HI
line observations which could in principle be done at no extra cost in
terms of telescope time. The rotation velocity technique shares many
of the characteristics of the polarization approach described above --
in the limit of perfectly well-behaved disk galaxies, it is also free
of shape noise and it can also be used to remove the contaminating
effect of intrinsic galaxy alignments. In practice, the degree to
which the rotational velocity technique improves on standard methods
will be dependent on observational parameters analogous to the ones
for polarization discussed above. First, one would need to account for
the fact that the HI line emission of galaxies is much fainter than
the broad-band continuum emission and so the number of galaxies will
be reduced (equivalent to the $n_{\rm pol}$ parameter discussed
above); and secondly, for a population of real disk galaxies, there
will again be some scatter in the relationship between the rotation
axis and the major axis of the galaxy disk (equivalent to the
$\alpha_{\rm rms}$ parameter in the polarization case). Recently,
\cite{huff13} have proposed to extend this technique using the
Tully-Fisher relation to calibrate the rotational velocity shear
measurements and thus reduce the residual shape noise even further. 

Both the polarization technique and the rotation velocity approach are
currently being tested as part of the SuperCLASS and CHILES
projects. They offer great promise for reducing the impact of shape
noise and intrinsic alignments in radio weak lensing surveys. We note
that the rotation velocity approach would not be feasible with a
L-band A-array configuration survey due to the low surface brightness
in HI. Nevertheless, the application of the polarization approach on
the VLASS data is likely to be one of the most exciting aspects of the
radio weak lensing analysis.

In addition to these new astrophysical probes, radio observations
offer unique advantages for traditional lensing analyses by way of
fitting for galaxy shapes directly from $uv$-visibility
data~\citep{chang02, chang04}. Another unique aspect comes from the
JVLA's wide bandwidth at L- and S-band which will allow the direct
measurement of the frequency dependence of the beam. This is a
potential major advantage over optical broad-band photometry where
galaxy SEDs vary wildly while in the radio, galaxies typically exhibit
smooth power-law type spectra.

\section{Choice of observing fields and overlap with optical surveys}
\label{observing_fields}

In Section~\ref{survey_optimize} we have identified deep,
high-resolution (A-array) L-band observations of relatively small sky
areas ($\sim 10-100$ square degrees) as the optimal survey
configuration from the point of view of maximising the sensitivity for
cosmological weak lensing measurements. If the VLASS is to pursue such
a strategy, the choice of field location is obviously a key
consideration. Here, we identify the regions on the sky that are
accessible to the JVLA and for which deep lensing quality optical and
near infra-red (NIR) imaging already exists. The deep NIR data will be
crucial as it will allow the estimation of photometric redshifts for
galaxies in the range $1 < z < 2$ where a large fraction of the VLASS
source galaxies are predicted to lie (see Fig.~\ref{fig:nofz_dist}).

\begin{figure}[t!]
\vspace{-5mm}
\centering
\includegraphics[width=15cm]{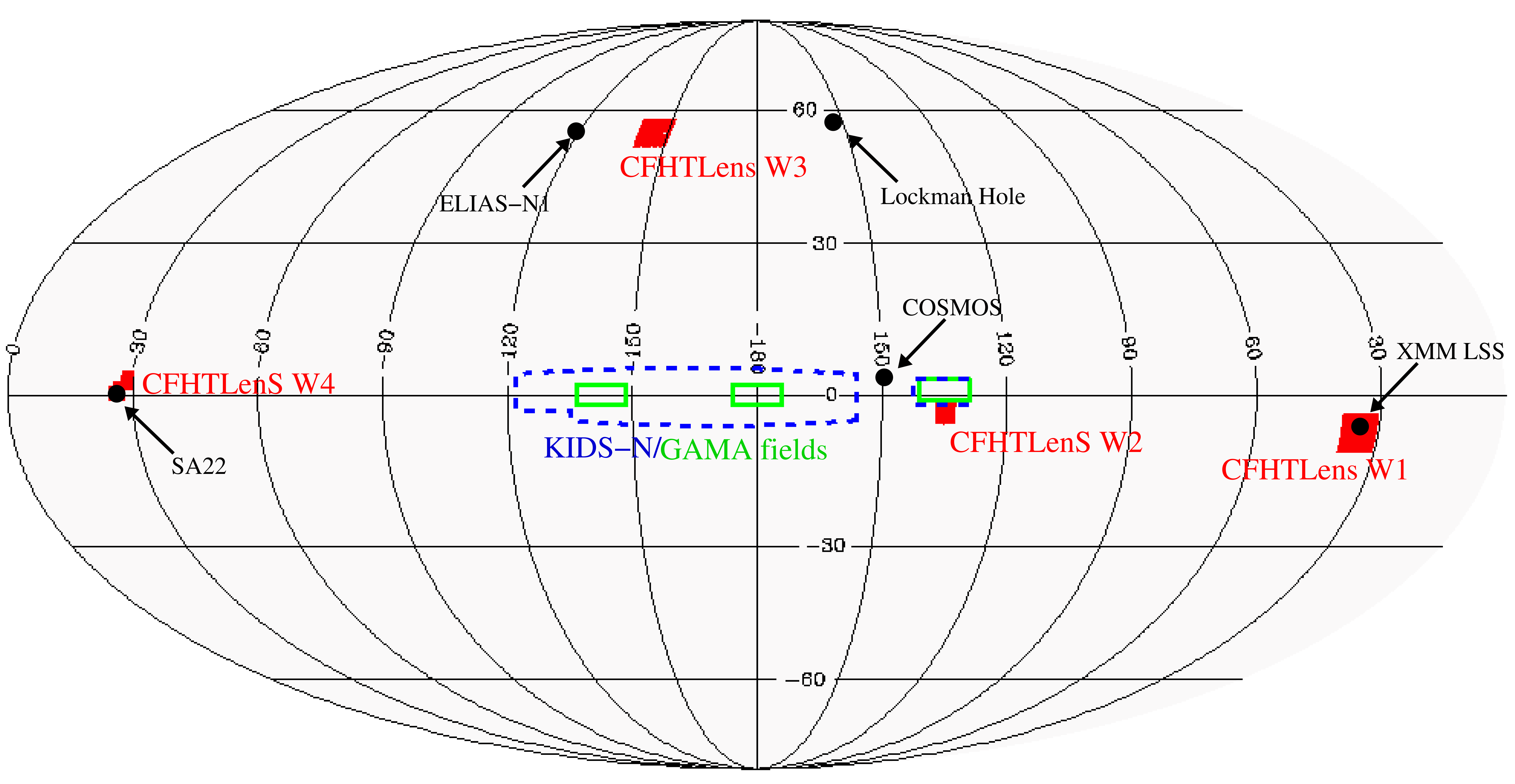}
\vspace{1mm}
\caption{Field locations for some of the major deep optical and NIR
  surveys at Dec $\gtrsim 0^\circ$. The projection is centred on
  (R.~A.~\!, Dec.) = (12 hrs, 0$^\circ$). The four CFHTLenS fields are shown
  along with the large-area KiDS-N/GAMA field. The five fields
  containing deep NIR imaging crucial for obtaining redshift estimates
  for the proposed VLASS observations are indicated with filled
  black circles. Together they comprise $\sim\!\!10$--20 square degrees,
  ideally matched to the L-band VLASS survey area preferred in the
  lensing optimization analysis described in
  Section~\ref{survey_optimize}. These five well-studied regions would
  be excellent candidates for a program of targeted deep VLASS
  fields.}
\label{fig:survey_fields}
\end{figure}

Fig.~\ref{fig:survey_fields} shows the location of major northern and
equatorial survey fields where deep optical and NIR imaging already
exists. In addition to the CFHTLenS and KiDS surveys (which provide
high-quality optical imaging data over large areas), also displayed
are the location of five key deep field -- the XMM-LSS, the Lockman
Hole, the ELAIS N1 field, the SA22 field and the COSMOS field. These
areas all include deep NIR observations (e.g. from the UKIDSS-DXS or
VISTA VIDEO surveys) and hence would be ideal for
supplying photometric redshift information to a VLASS deep fields
survey program. Each also offers a host of complementary observations
at other wavebands.

In addition to providing redshifts, overlapping high quality optical
imaging in these fields will provide a unique opportunity to test
cross-correlation techniques that have been proposed with a view to
mitigating instrumental systematics in weak lensing
analyses~\citep{jarvis08, patel10}. Although not the focus of this
white paper, we note in passing that a deep A-array L-band survey
targeting these well-studied fields would prove very interesting for a
host of other extra-galactic science areas such as galaxy formation
and evolution, star-formation studies out to high redshift, galaxy
morphology studies and 3D clustering analyses of AGN and starburst
galaxies.
 
\section{Conclusions}
\label{conclusions}
We have explored the potential of large JVLA survey programs 
to provide a major step forward in the field of radio
weak lensing. Of all of the SKA precursor and pathfinder telescopes, the
JVLA is perhaps the instrument that is most suited to weak lensing
work thanks to its unique combination of excellent sensitivity,
relatively high angular resolution and relatively fast survey
speeds. This white paper argues for a deep and high-resolution survey
conducted over a relatively small survey area (a few 10s of square
degrees) at L-band in A-array configuration. Such a survey will build
on the JVLA's key strengths and will enable ground-breaking radio weak
lensing science as well as many other compelling science goals in the
areas of galaxy evolution, AGN clustering and understanding
the physics and morphologies of star-forming and starburst galaxies to
high redshift. Our key findings are:

\begin{itemize} 
\item Depending on the duration of time devoted to a large VLASS
  program, a radio weak lensing analysis of the VLASS could detect the
  effects of weak lensing by large scale structure with significances
  up to $\sim10\sigma$. To achieve the upper limit of
  this range would require a 10000 hr deep radio survey over
  $\sim10$--20 deg$^2$ to be conducted at L-band in A-array
  configuration. 
\item Up to $7\sigma$ ($5\sigma$) detections could be achieved with
  a 5000 hr (3000 hr) L-band survey. The optimal survey area would be
  to focus on deep fields covering $\sim$10 square degrees. 
\item The weak lensing potential of the VLASS is reduced with
  increasing frequency. S-band could achieve a $4\sigma$ detection
  with a 5000 hr survey while a survey conducted at C-band would
  only exceed a $3\sigma$ detection with 10000 hrs of telescope
  time. 
\item The radio band offers unique advantages for performing weak
  lensing studies. Both polarization information and rotational
  velocity measurements from HI line surveys hold great promise for
  reducing the effects of shape noise and minimizing the contaminating
  effects of intrinsic alignments. The polarization technique can only
  be used in the radio. 
\item If the VLASS are to pursue a deep fields strategy, the choice
  of fields should be informed by the location of existing
  high-quality optical and NIR data. We have identified five
  fields covering $\sim\!\!10$--20 square degrees where this information
  already exists and which would be good candidates for a VLASS deep
  fields program.  
\end{itemize}

\begin{figure}[t!]
\vspace{-10mm}
\centering
\includegraphics[width=12cm]{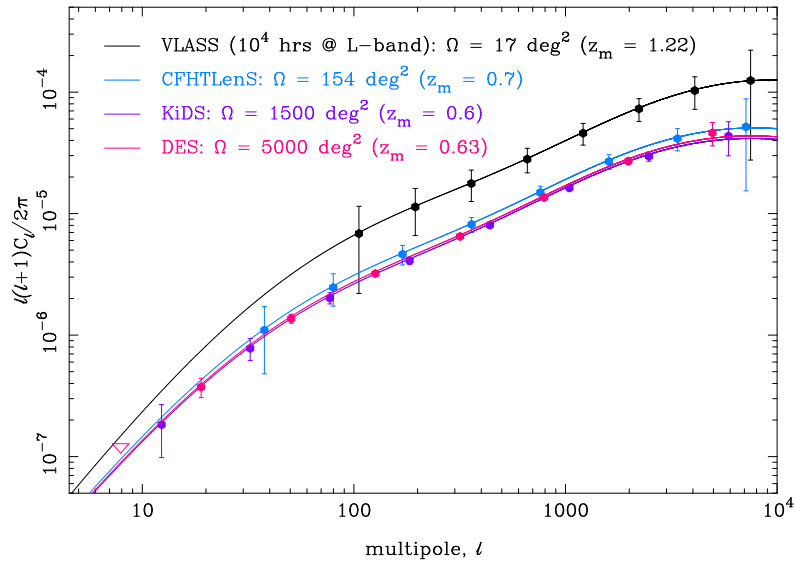}
\vspace{-5mm}
\caption{Performance of the optimal 17 deg$^2$ 10000-hr L-band survey
  compared to the precision of current state-of-the art optical weak
  lensing survey (CFHTLenS) and two of the major ongoing weak lensing surveys
  (KiDS and DES). The KiDS, DES and HSC surveys will be conducted over the
  next 5 years, potentially on a similar timescale to a major
  VLASS. Note how the VLASS probes a much higher redshift range than
  the optical lensing surveys. The proposed VLASS would thus provide
  additional cosmological information on the evolution of structure at
  high-$z$ beyond the reach of the optical-based lensing surveys.
  The forecasted detection significances of the constraints
  shown are $\sim10\sigma$ (VLASS), $15\sigma$ (CFHTLenS),
  $30\sigma$ (KiDS) and $43\sigma$ (DES).}
\label{fig:compare_optical}
\end{figure}

We conclude with a demonstration of how a weak lensing analysis of a
VLASS deep fields program could significantly enhance the current
state-of-the-art in terms of optical weak lensing
measurements. Fig.~\ref{fig:compare_optical} shows the forecasted
constraints from the proposed $\sim17$ deg$^2$ 10000 hr L-band survey
with the corresponding forecasts for the current state-of-the-art
survey (CFHTLenS) as well as the forecasts for two representative
ongoing optical surveys, the KiDS and DES surveys. The latter two
surveys, and the HSC survey, are to be conducted over the next 5
years, potentially on a similar timescale to a large VLASS program. We
immediately see that the VLASS survey probes a redshift range that is
completely complementary to the shallower redshifts of the optical
surveys. The addition of the high redshift information from the VLASS
would therefore greatly enhance studies of the growth of structure
through lensing by adding additional high-redshift bins to a
tomographic cosmic shear analysis. This would in turn result in
improved constraints on dark energy parameters. Beyond this, novel
radio-based approaches to weak lensing through the use of polarization
and rotational velocity measurements may well yield the best route
forward for dealing with the problem of intrinsic galaxy alignments,
the most troublesome obstacle for future precision cosmology
experiments based on weak lensing. Designing the forthcoming VLASS to
accommodate the weak lensing science exploitation advocated in this
paper will provide an ideal dataset on which to demonstrate and refine
these novel techniques.

\small
\bibliographystyle{mn2e}
\bibliography{vlass_weak_lensing}
\vspace{0.5cm}
\hrule
\vspace{0.5cm}
\noindent
 $^{1}$Jodrell Bank Centre for Astrophysics, University of Manchester, Oxford Road, Manchester M13 9PL, UK\\ 
 $^{2}$Department of Physics and Astronomy, University College London, Gower Street, London WC1E 6BT, UK\\
 $^{3}$Institute for Astronomy, ETH Zurich, Wolfgang-Pauli-Strasse 27, CH-8093 Zurich, Switzerland\\
 $^{4}$Institute for Cosmology \& Gravitation, University of Portsmouth, Portsmouth, PO1 3FX, UK\\
 $^{5}$Max-Planck-Institut f\"ur Astrophysik, Karl-Schwarzschildstr.~1, 85748 Garching, Germany\\
 $^{6}$HH Wills Physics Laboratory, University of Bristol, Tyndall Avenue, Bristol BS8 1TL, UK\\
 $^{7}$Department of Physics \& Astronomy, University of California, Irvine, CA 92697, USA\\
 $^{8}$Ludwig-Maximilians-Universit\"at M\"unchen, Geschwister-Scholl-Platz 1, 80539 Munich, Germany\\
 $^{9}$Astrophysics, University of Oxford, Denys Wilkinson Building, Keble Road, Oxford, OX1 3RH, UK\\ 
$^{10}$School of Physics \& Astronomy, University of Nottingham, University Park, Nottingham NG9 2RD, UK\\
$^{11}$Janksy Fellow; National Radio Astronomy Observatory, Socorro, NM, 87801, USA\\
$^{12}$Imperial Centre for Inference \& Cosmology, Imperial College, Blackett Lab., London, SW7 2AZ, UK\\
$^{13}$Institute for Astronomy, University of Edinburgh, Blackford Hill, Edinburgh EH9 3HJ, UK\\ 
$^{14}$Institute for Astronomy, University of Hawaii, 2680 Woodlawn Drive, Honolulu, HI 96822, USA\\
$^{15}$Mullard Space Science Laboratory, UCL, Holmbury St Mary, Dorking, Surrey RH5 6NT, UK\\
$^{16}$Department of Physics, University of the Western Cape, Bellville 7535, South Africa\\
$^{17}$National Radio Astronomy Observatory, Socorro, NM, 87801, USA\\
$^{18}$Astrophysics, Cosmology Gravity Centre, University of Cape Town, Cape Town, 7701, South Africa\\
$^{19}$Jet Propulsion Laboratory, California Institute of Technology, Pasadena CA 91109, USA \\
$^{20}$California Institute of Technology, Pasadena CA 91125, USA\\
$^{21}$School of Physics \& Astronomy, University of Southampton, Highfield, Southampton, SO17 1BJ, UK\\
$^{22}$CENTRA, IST, Universidade Tecnica de Lisboa, Av. Rovisco Pais 1, 1049-001 Lisboa, Portugal\\
$^{23}$Universit\"at Heidelberg, Monchhofstra\ss e 12, 69120 Heidelberg, Germany\\
$^{24}$Insititute for Computational Cosmology, Durham University, South Road, Durham DH1 3LE, UK\\
$^{25}$CEA Saclay, IRFU, Service d'Astrophysique, 91191 Gif-Sur-Yvette CEDEX, France\\

\end{document}